# Supporting Musical Practice Sessions Through HMD-Based Augmented Reality

**Karola Marky**
Technische Universität Darmstadt
Darmstadt, Germany
marky@tk.tu-darmstadt.de

**Andreas Weiß**
Musikschule Schallkultur
Kaiserslautern, Germany
andreas.weiss@musikschule-schallkultur.de

**Thomas Kosch**
LMU Munich
Munich, Germany
thomas.kosch@ifi.lmu.de

## ABSTRACT

Learning a musical instrument requires a lot of practice, which ideally, should be done every day. During practice sessions, students are on their own in the overwhelming majority of the time, but access to experts that support students "just-in-time" is limited. Therefore, students commonly do not receive any feedback during their practice sessions. Adequate feedback, especially for beginners, is highly important for three particular reasons: (1) preventing the acquirement of wrong motions, (2) avoiding frustration due to a steep learning curve, and (3) potential health problems that arise from straining muscles or joints harmfully. In this paper, we envision the usage of head-mounted displays as assistance modality to support musical instrument learning. We propose a modular concept for several assistance modes to help students during their practice sessions. Finally, we discuss hardware requirements and implementations to realize the proposed concepts.

## 1 INTRODUCTION AND BACKGROUND

When learning a musical instrument, beginners need to acquire a range of different skills. They have to learn constructional details and the music notation that is specific for each musical instrument. Furthermore, students have to learn how to interact correctly with the musical instrument in order to play it. This includes the knowledge of the right movements and postures for different body parts. The transformation of this knowledge into motions and the development of muscle enhance the memorization of the played musical piece.

The acquirement of knowledge and its transformation requires a lot of deliberate practice and patience [6, 12, 23]. Thus, during deliberate practice, the students strive to improve their knowledge and skill set and practice by themselves in the overwhelming majority of the time. Traditional supportive materials, such as textbooks or videos, lack of feedback options. The guidance by experts provides only a temporal solution because experts are costly and not available on-demand. However, especially beginners require adequate feedback during deliberate practice for several reasons. For example, beginners might learn wrong movements or postures and the correction of those is difficult and time-consuming later on. Furthermore, excessive training of falsely acquired movements may lead to serious injuries [19]. Moreover, the overall focus may decrease during times of over-practicing which may lead to frustration due to absent progress. Hence, students – especially beginners – should be supported during deliberate practice with adequate feedback to avoid the issues mentioned above.

A plethora of technology-based learning setups and concepts have been proposed in the literature or is available by commercial vendors. All of them use different techniques either to extend traditional musical instruments or to transform them into a smart object. Additional components, such as projectors [13], sensors [2], or actuators [8, 25] can be mounted on the musical instrument. Furthermore, lights [5, 7, 22, 24] or sensors [22] can be integrated in it to provide visible feedback. Furthermore, Augmented Reality (AR) based on mobile devices can provide assistance [4].

However, musical instruments are costly and it is rather unlikely that beginners purchase a musical instrument that includes an assistance system for learning purposes. Mounting additional components is similar to having a specific instrument: the students would have to acquire them, or the schools would have to loan them. Techniques that are based on tracking or projecting could require a special setup that has to be mounted at the home of the student. Furthermore, existing approaches are affected by several drawbacks [15]. First, they are mainly focused on the positioning of finger targets. They lack posture information and focus mainly on where the students have to place their fingers. Second, they are unresponsive and cannot adjust to the student's current needs in real-time. This includes the adjustment of waiting times in which the student has to put their hands into place. Third, they might interfere with playing the instrument or distract the students. For instance, by incorporating markers or sensors on the instrument that get in the way, the learning experience can be disrupted.

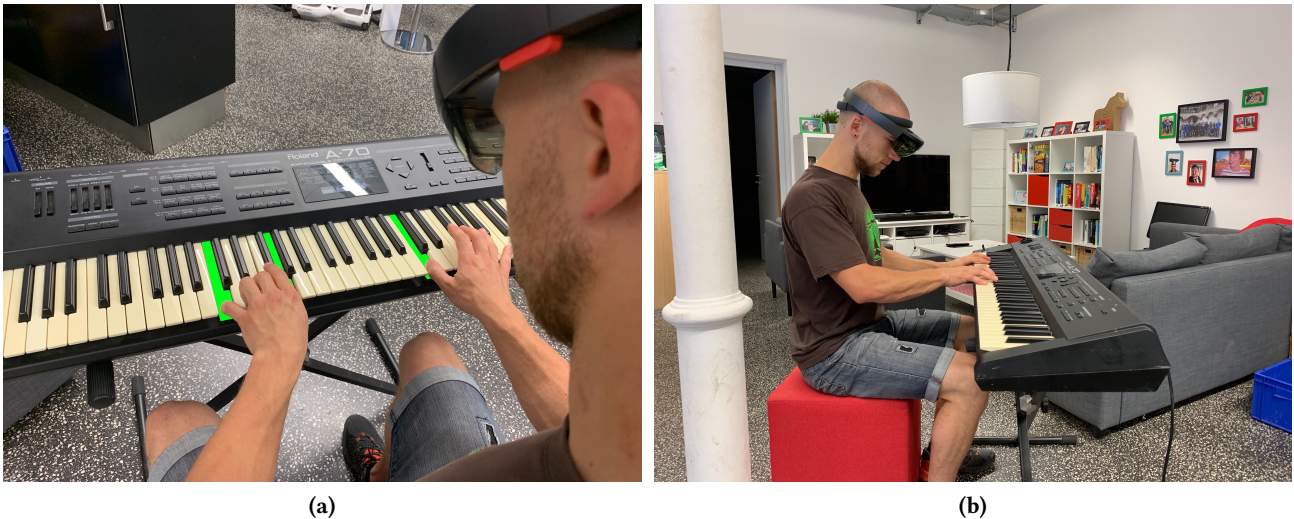

(a)  (b)

**Figure 1: Student using an HMD to practice piano. HMDs have the potential to provide visual and auditory assistance during practice sessions.**

We argue that feedback by an AR setup with an HMD is a potential solution for providing students with adequate feedback during practice, since AR technologies hold promise to support an independent learning and education processes [10, 11]. To provide adequate feedback to the students and to overcome the described drawbacks, we envision the usage of an HMD with visual and auditory AR cues (see Figure 1). Current technologies embedded in HMDs are capable of assessing the students' movements and postures. In the remainder of this paper, we describe our proposed concept and its features. To support the design and feasibility of our envisioned concept, we provide explanations that are based on existing user studies in the literature. We conclude with an outlook for future work.

## 2  LEARNING INSTRUMENTS THROUGH AR

AR setups that need to project mobile content in front of the user are required to be mounted on the upper forehead. We envision those setups as mobile companions that connect and interact with nearby devices. For example, the interaction space can be extended by including smartphones or smartwatches. Furthermore, the assessment of the musical performance can be reflected afterward on these devices. HMDs are suitable for providing real-time assistance as the content is displayed immediately to support the student. In this section, we describe features as well as benefits that our musical AR learning concept provides and how they can be implemented. Hereby, we focus on string, key, and percussion instruments[1]. Figure 2 illustrates the modalities with their according support modes which we envision to incorporate into an HMD application.

**Cognitive Assistance**

Learning a musical instrument requires a combination of theoretical and practical skills. Theoretical skills include remembering the arrangement of notes which vary with each instrument. Furthermore, reading notes requires to map written note representations to their auditory counterpart and position on the musical instrument in real-time. This mapping increases the perceived cognitive workload while playing a new instrument without suitable difficulty adaption. Past research used physiological sensing as an indirect measure to change the difficulty in real-time [26, 27]. However, this requires to apply cumbersome sensors on the body. In the following, we discuss how we envision HMDs as musical practice assistants that provide cognitive alleviation through in-situ feedback without the need for external sensors.

---

[1] The interaction with wind instruments is commonly outside the student's field of view and the recognition of blowing requires special equipment which is not in the scope of this paper.

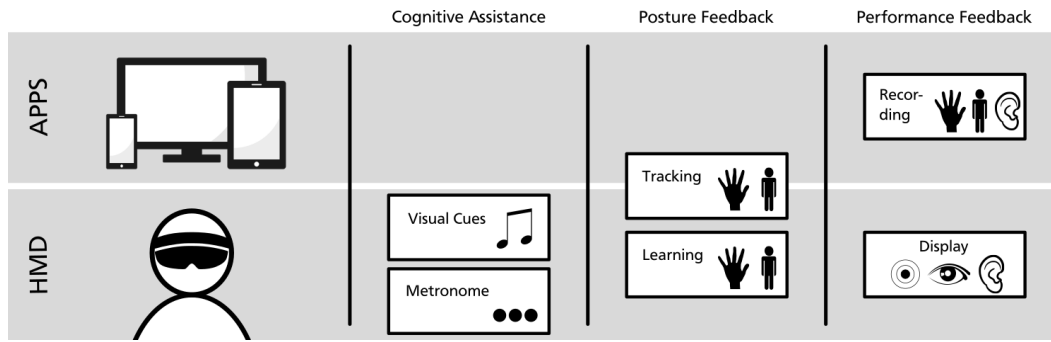

Figure 2: Overview of our proposed features and modes that are included in an HMD application.

*Visual Cues Mode.* Beginners have to understand and interpret the music notation which is different among musical instruments. Playing, based on this notation, is difficult for beginners because the information is on a separate source, such as a sheet of paper. Students have to map it from the source to the instrument in real-time to enable a fluent musical experience. Furthermore, the students have to remember how this particular information is mapped to the individual instrument, hence increasing the overall perceived cognitive workload. To unload this mental demand from beginners, our setup provides the following feature that is inspired by previous research [3, 13, 14, 18, 20]. Part of our proposed concept is the visualization of cues on the instrument through AR. Hereby, visual target markers are displayed directly on the surface of the instrument. The targets can have different colors and each color corresponds to one finger. These targets help the students to place their fingers without having to perform a mapping and therefore alleviate cognitive workload. In a study of a guitar with a fretboard that integrates LEDs [9] it has been shown, that finger targets on the fretboard can enhance the performance of beginners. This feature does not substitute the learning of the music notation, but assists the student and enables quick access to the musical instrument.

It has been shown that the cognitive load of beginning piano players can be reduced by a depicting the next keys as vertical bars that flow from the top towards a key [3]. Therefore, this representation of a piano song can be chosen in our concept.

*Metronome Mode.* A metronome is a device that assists the student in learning to play to a regular beat. To do so, the metronome produces a sound at a certain tempo, typically in beats per minute, that is set by the student. Our envisioned concept contains a metronome mode that can be switched on and adjusted by the student. The metronome can be played auditory only through an HMD or be displayed as visible dots in the student's field of view.

**Posture Feedback**

The postures of fingers as well as the body, are of great importance during practice. For instance, if the fingers are overstretched or too cramped, the student applies more force than required. Overstretched joints also reduce the student's accuracy. Wrong postures can even lead to health considerations, such as the repetitive strain injury [21] or sore body parts. Therefore, we envision detailed body and finger information through HMD-based tracking. Bad postures can then be detected and communicated with the student to improve the overall body and finger positioning. In the following, we provide details on posture-related features that can be incorporated into HMDs. Feedback can be provided in real-time or as a visualization that is displayed after a training session to increase self-awareness.

*Learning Mode.* The student can choose a mode for learning new and improving known postures. In this mode, two types of postures can be learned: (1) finger and hand postures as well as (2) body postures. In the first case, a 3D-model of the correct finger or hand posture is displayed by the HMD [17]. Students use the 3D-model as a visual guide and place their hands accordingly to receive visual or haptic feedback when they succeed. The visual feedback can be an optical cue in the student's field of view. The haptic feedback can be a vibration of the HMD. For the second case, a gyroscope inside the HMD tracks the balance of the student [1]. The students receive visual feedback about their balance and visual cues on how to adjust their body posture. This feedback can be a 3D-model of a person in the student's field of view.

Note, that the learning mode is not used during active playing, its purpose is to raise the students' awareness for their postures and to develop knowledge about correct postures.

*Tracking Mode.* In a second posture-related mode, the students' postures are tracked during active playing. If the students perform postures incorrectly, they can either receive visual or haptic feedback, depending on their personal preference. The visual feedback can be a highlight of the area with the incorrect position or an arrow that highlights the direction in which the student should optimally move.

If the visual channel requires the full attention of the student, haptic feedback can be employed as an alternative. If a harmful body posture is detected, vibrations at the according part of the upper head can be provided to indicate a movement in this direction. Note, that the feedback can be either immediate during active playing, or after the student has finished playing.

**Performance Feedback**

Performance feedback can be provided to reflect on past learning progress. In the following, we outline several modes that implement our vision.

*Display Mode.* In this mode, students can get feedback to an actively played musical piece. The student can choose between haptic, visual, or auditory feedback. Before playing, the students pick a sequence of chords or tones and a rhythm. While playing, the HMD tracks the students' movements and provides feedback of incorrectly played chords or tones.

*Recording Mode.* In the recording mode, the performance of the student is recorded. Later the performance the student can play a visual representation of his or her performance together with a previously loaded reference piece of music.

## 3 INTERACTION AND HARDWARE REQUIREMENTS

The implementation of our proposed setup requires mobile hardware and a user-friendly interface. We show that a prototype can already be built on existing hardware and provide guidance for designing a user interface on a complementary second device.

**Hardware**

With the progression of technology, AR setups are becoming smaller and less tedious to use. For instance, the Microsoft HoloLens 2 [2] provides tracking of individual fingers without additional hardware. Although nowadays AR glasses are quite heavy, with the progress of technological innovation they will become lighter in the future.

**User Interfaces**

A user interface is required to adjust different practical lessons and modes. Without additional devices, most HMDs offer a set of gestures that can be used to configure and select the outlined modes. However, this limits the interaction space to gestures only. Furthermore, most HMDs possess a narrow field of view which limits the amount of information that can be processed. Previous research showed the potential of using desktop and smartphone applications [16] to control and display complex information. Thus, we envision that HMDs are suitable to display simplified feedback in real-time while adjusting and reviewing past practice sessions on complementary devices with larger screen space.

**Combining Modes**

The envisioned setup enables a combination of the presented modes. Combinations of modes may increase the overall assistance or enable the student to play more challenging musical pieces. For example, the *visual cues mode* and *metronome mode* may support fluent playing while keeping the beat. Combining the *learning mode* and the *tracking mode* enable students to reflect on their practice performance. Possible modes combinations and their effectiveness form an integral part of future work analogously to the interface design mentioned above.

## 4 CONCLUSION AND OUTLOOK

In this paper, we propose a modular AR concept that is based on an HMD and a mobile device for supporting students during deliberate practice. Our concept provides different modes that can be chosen and adjusted by the students according to their needs. The students can either use the provided realtime assistance or reflect their musical performance afterward. In the first step, the feasibility and user acceptance of the presented prototype application will be assessed within qualitative inquiries. This enables us to receive insights into desired functionalities by music teachers and students. Afterward, the emerged functionalities will be implemented and evaluated in a user study. The necessary technology and hardware requirements are already available within a single device. Therefore, we believe in the flexible usage of HMDs which provide a personal trainer for beginners that can be independently used regarding the learning progress or used musical instrument.

## REFERENCES

[1] Oliver Amft, Florian Wahl, Shoya Ishimaru, and Kai Kunze. 2015. Making Regular Eyeglasses Smart. *IEEE Pervasive Computing* 14, 3 (2015), 32–43. https://doi.org/10.1109/MPRV.2015.60

[2] Frédéric Bevilacqua, Nicolas H. Rasamimanana, Emmanuel Fléty, Serge Lemouton, and Florence Baschet. 2006. The Augmented Violin Project: Research, Composition and Performance Report. In *Proceedings of the 6th International Conference on New Interfaces for Musical Expression (NIME 06)*. HAL, 402–406.

[3] Jonathan Chow, Haoyang Feng, Robert Amor, and Burkhard C. Wünsche. 2013. Music Education Using Augmented Reality with a Head Mounted Display. In *Proceedings of the Fourteenth Australasian User Interface Conference (AUIC '13)*. Australian Computer Society, Inc., Darlinghurst, Australia, Australia, 73–79. http://dl.acm.org/citation.cfm?id=2525493.2525501

[4] Shantanu Das, Seth Glickman, Fu Yen Hsiao, and Byunghwan Lee. 2017. Music Everywhere – Augmented Reality Piano Improvisation

---

[2]www.microsoft.com/en-us/hololens/hardware – last access 2019-07-12

Learning System. (2017).
[5] LLC. Edge Tech Labs. 2019. FretZealot. http://fretzealot.com/. [Online; accessed: 14-March-2019].
[6] K. Anders Ericsson, Ralf T. Krampe, and Clemens Tesch-Römer. 1993. The Role of Deliberate Practice in the Acquisition of Expert Performance. *Psychological Review* 100, 3 (1993), 363.
[7] CASIO America Inc. 2019. Casio LK-280. https://www.casio.com/products/electronic-musical-instruments/lighted-keys/lk-280. [Online; accessed: 14-March-2019].
[8] Jakob Karolus, Hendrik Schuff, Thomas Kosch, Paweł W. Wozniak, and Albrecht Schmidt. 2018. EMGuitar: Assisting Guitar Playing with Electromyography. In *Proceedings of the Designing Interactive Systems Conference (DIS '18)*. ACM, New York, NY, USA, 651–655.
[9] Joseph R. Keebler, Travis J. Wiltshire, Dustin C. Smith, and Stephen M. Fiore. 2013. Picking up STEAM: Educational Implications for Teaching with an Augmented Reality Guitar Learning System. In *International Conference on Virtual, Augmented and Mixed Reality*. Springer, Cham, Switzerland, 170–178.
[10] Pascal Knierim, Thomas Kosch, Matthias Hoppe, and Albrecht Schmidt. 2018. Challenges and Opportunities of Mixed Reality Systems in Education. In *Mensch und Computer 2018 - Workshopband*. Gesellschaft für Informatik e.V., Bonn, 325–330.
[11] Thomas Kosch, Pascal Knierim, Paweł Woźniak, and Albrecht Schmidt. 2017. Chances and Challenges of Using Assistive Systems in Education. In *Mensch und Computer 2017 - Workshopband*. Gesellschaft für Informatik e.V., Bonn, 389–394.
[12] Ralf Th. Krampe and Karl A. Ericsson. 1995. *Deliberate Practice and Elite Musical Performance.* Cambridge University Press Cambridge, United Kingdom, 84–102.
[13] Markus Löchtefeld, Sven Gehring, Ralf Jung, and Antonio Krüger. 2011. guitAR: Supporting Guitar Learning Through Mobile Projection. In *Proceedings of the Extended Abstracts on Human Factors in Computing Systems (CHI EA '11)*. ACM, New York, NY, USA, 1447–1452.
[14] Markus Löchtefeld, Sven Gehring, Ralf Jung, and Antonio Krüger. 2011. Using Mobile Projection to Support Guitar Learning. In *Proceedings of the International Symposium on Smart Graphics*. Springer, Cham, Switzerland, 103–114.
[15] Karola Marky, Andreas Weiß, Julien Gedeon, and Sebastian Günther. 2019. Mastering Music Instruments through Technology in Solo Learning Sessions. In *Proceedings of the 7th Workshop on Interacting with Smart Objects (SmartObjects '19)*. CEUR Workshop Proceedings, 1–7.
[16] Alexandre Millette and Michael J. McGuffin. 2016. DualCAD: Integrating Augmented Reality with a Desktop GUI and Smartphone Interaction. In *In Proceedings of the IEEE International Symposium on Mixed and Augmented Reality (ISMAR-Adjunct)*. IEEE, 21–26. https://doi.org/10.1109/ISMAR-Adjunct.2016.0030
[17] Yoichi Motokawa and Hideo Saito. 2006. Support System for Guitar Playing Using Augmented Reality Display. In *Proceedings of the 5th IEEE and ACM International Symposium on Mixed and Augmented Reality (ISMAR '06)*. IEEE, Washington, DC, USA, 243–244.
[18] Honghu Pan, Xingxi He, Hong Zeng, Jia Zhou, and Sai Tang. 2018. Pilot Study of Piano Learning with AR Smart Glasses Considering Both Single and Paired Play. In *Proceedings of the International Conference on Human Aspects of IT for the Aged Population (ITAP 2018)*. Springer, 561–570.
[19] A. B. M. (Boni) Rietveld. 2013. Dancers' and musicians' injuries. *Clinical Rheumatology* 32, 4 (01 Apr 2013), 425–434.
[20] Katja Rogers, Amrei Röhlig, Matthias Weing, Jan Gugenheimer, Bastian Könings, Melina Klepsch, Florian Schaub, Enrico Rukzio, Tina Seufert, and Michael Weber. 2014. P.I.A.N.O.: Faster Piano Learning with Interactive Projection. In *Proceedings of the International Conference on Interactive Tabletops and Surfaces (ITS '14)*. ACM, New York, NY, USA, 149–158.
[21] M. Rosety-Rodriguez, F. J. Ordóñez, J. Farias, M. Rosety, C. Carrasco, A. Ribelles, J. M. Rosety, and M. Gomez Del Valle. 2019. The Influence of the Active Range of Movement of Pianists' Wrists on Repetitive Strain Injury. *European Journal of Anatomy* 7, 2 (2019), 75–77.
[22] Yejin Shin, Jemin Hwang, Jeonghyeok Park, and Soonuk Seol. 2018. Real-time Recognition of Guitar Performance Using Two Sensor Groups for Interactive Lesson. In *Proceedings of the Twelfth International Conference on Tangible, Embedded, and Embodied Interaction (TEI '18)*. ACM, New York, NY, USA, 435–442. https://doi.org/10.1145/3173225.3173235
[23] John A. Sloboda, Jane W. Davidson, Michael J.A. Howe, and Derek G. Moore. 1996. The Role of Practice in the Development of Performing Musicians. *British Journal of Psychology* 87, 2 (1996), 287–309.
[24] The Fretlight Guitar Store. 2019. FretLight. https://fretlight.com/. [Online; accessed: 14-March-2019].
[25] Janet van der Linden, Erwin Schoonderwaldt, Jon Bird, and Rose Johnson. 2011. MusicJacket – Combining Motion Capture and Vibrotactile Feedback to Teach Violin Bowing. *IEEE Transactions on Instrumentation and Measurement* 60, 1 (Jan 2011), 104–113.
[26] Beste F. Yuksel, Daniel Afergan, Evan M. Peck, Garth Griffin, Lane Harrison, Nick WB Chen, Remco Chang, and Robert J.K. Jacob. 2015. Braahms: a Novel Adaptive Musical Interface Based on Users' Cognitive State. In *Proceedings of the Conference on New Interfaces for Musical Expression (NIME '15)*. 136–139.
[27] Beste F. Yuksel, Kurt B. Oleson, Lane Harrison, Evan M. Peck, Daniel Afergan, Remco Chang, and Robert JK Jacob. 2016. Learn Piano with BACh: An Adaptive Learning Interface That Adjusts Task Difficulty Based on Brain State. In *Proceedings of the 2016 CHI Conference on Human Factors in Computing Systems (CHI '16)*. ACM, New York, NY, USA, 5372–5384. https://doi.org/10.1145/2858036.2858388